\documentclass[a4paper,12pt]{article}
\usepackage[dvips]{graphicx}
\usepackage{braket}
\usepackage{bbm}
\usepackage{amssymb}
\usepackage{amsmath}
\usepackage{amsthm}

\DeclareMathOperator{\Tr}{Tr}
\DeclareMathOperator{\Prob}{Prob}

\begin{document}

%==========================================================================================
% TITLE PAGE
%==========================================================================================

\begin{center}
\noindent{\large \bf Reversibility and Irreversibility within the
Quantum Formalism
\normalsize}\\
\vspace{10pt} {\bf Tim Jacobs} %\footnote{corresponding author: {\tt tim.jacobs@fys.kuleuven.ac.be}}
and {\bf Christian Maes} \\
{Instituut voor Theoretische Fysica\\ K.U.Leuven, B-3001 Leuven, Belgium.} \\ %\vspace{6pt}\\
%\today\\
\end{center}

%\newpage
~\vspace{0,5cm}~

%==========================================================================================
\subsection*{Setting}
%==========================================================================================

The discussion on time-reversal in quantum mechanics exists at
least since Wigner's paper \cite{Wigner} in 1932. If and how the
dynamics of the quantum world is time-reversible has been the
subject of many controversies. Some have seen quantum mechanics as
fundamentally time-irreversible, see for example von Neumann
\cite[p.\,358]{vonNeumann}, and some have seen in that the
ultimate cause of time's arrow and  second law behavior. In his
best-selling book, Roger Penrose argues similarly and concludes
that ``\textit{our sought-for quantum gravity must be a
time-asymmetric theory}'' \cite[p.\,351]{Penrose}. Not so long ago,
we read yet about another project in Physicalia, \cite{deHaan}: to extend quantum mechanics into new fundamentally
irreversible equations, thus proposing a new theory giving
``\textit{... une description fondamentale irr\'{e}versible de
tout syst\`{e}me physique}''. \\

We take here  the opportunity to state and review a number of
general points that are perhaps less emphasized in the existing
literature. We first explain what is meant by mechanical
reversibility and how it applies in classical mechanics and,
somewhat differently, for the free evolution in quantum mechanics.
If the evolution is not free, i.e., it is interrupted by
measurements, the quantum formalism is challenged by the problem
of retrodiction.  Related to that is the notion of statistical
reversibility which is very similar to what is more commonly known
as the condition of detailed balance, at least for stochastic
processes describing the spatio-temporal fluctuations in
equilibrium. Finally, we describe the emergence of thermodynamic
irreversibility. Here there is little difference between the
classical and the quantum set-up even though one could have
thought that the presence of a discrete energy spectrum prohibits
dissipation for Hamiltonian evolutions.

%==========================================================================================
\subsection*{Mechanical reversibility}
%==========================================================================================

Consider a classical mechanical system going through a sequence of
positions and momenta $(q_0,p_0), (q_1,p_1) \ldots, (q_t,p_t)$.
That evolution solves Newton's equations of motion for given
forces, e.g. gravity.  Upon playing the movie backwards, i.e.,
time reversed, we see the system evolving from the positions $q_t$
to $q_0$, but now with reversed momenta $-p_t$ to $-p_0$. The
time-reversed sequence $(q_t,-p_t), \ldots, (q_0,-p_0)$ can or
cannot be a solution of the same equation of motion.  For say free
fall, the time-reversed sequence certainly solves the same
Newton's law of gravity; with friction or for the damped
oscillator, that time-reversal symmetry is not present. In other
words, symmetry with respect to time-reversal amounts to having
identical mechanical laws for prediction and for
retrodiction.\\

Generalizing to the more abstract idea of a dynamical system for
which the state $x_t$ at an arbitrary time $t$ is given in terms
of a flow $x_t = f_t(x_0)$ which is invertible, we say that the system is {\it
mechanically} (or also, {\it dynamically})  {\it reversible} if
there exists a transformation $\pi$ of states with $\pi^2=1$, for
which at all times
\begin{equation} \label{eq:kineticinversalCL}
\pi f_t \pi = f_t^{-1}
\end{equation}
The transformation $\pi$ is often called the {\it kinematical}
time-reversal. For classical Hamiltonian systems, $f_t$ should be
thought of as the Hamiltonian flow on the (micro)states $x=(q,p)$
given in terms of the positions $q$ and the momenta $p$ of all the
particles, and $\pi$ as the involution $\pi(q,p) = (q,-p)$.
Mechanical reversibility expresses that first evolving forward in
time and then changing all momenta gives the same state as first
changing all momenta and then going back in
time\footnote{Time-reversal in classical systems goes of course
beyond the equations of Newton or Hamilton. A recent application
of time-reversal in acoustics is found in \cite{Snipers}.}.

%==========================================================================================
\subsection*{Statistical reversibility I}
%==========================================================================================

At many instances our understanding of physical phenomena involves
statistical considerations\footnote{Even when God does not play
dice and also for the description of the classical world,
depending on the scale of the phenomena, stochastic dynamics
enter. They can be the result of pure modelling or they appear as
an effective or reduced dynamics.  Classical examples are the
Langevin description of Brownian motion, the Onsager-Machlup
description of fluctuating hydrodynamics and the stochastically
driven Navier-Stokes equation for turbulence. Since Pauli's work
\cite{Pauli}, stochastic processes are also obtained as the result
of quantum considerations, for example also under situations where
the Golden Rule applies. In fact, a great deal of so called
classical stochastic processes, like Glauber and Kawasaki
dynamics, find their origin in studies of quantum relaxation
processes. Moreover, practically all quantum processes for open
systems that are obtained under the weak coupling limit are just
standard Markov processes.}. We start by
explaining what is a time-reversible stochastic process.\\

Restricting ourselves to stationary Markov processes, the law of
the dynamics is given in terms of transition probabilities $p(x,s
\rightarrow x',t)$ to find $x'$ at time $t$, when the system is in
state $x$ at time $s$. To make it simple, we suppose discrete time
and a finite state space. We thus have a sequence or trajectory
$X_0,X_1,\ldots,X_t,\ldots$ of variables as sampled from a
probability law that specifies
\[
\Prob[X_t =x] = \rho(x),\quad  \mbox{ the stationary distribution}
\]
and the transition probabilities
\[
\Prob[X_t =x' | X_{t-1}=x, X_{t-2}= a_2,\ldots, X_0= a_{t}] \equiv p(x,x')
\]

We look at that stochastic dynamics in the time-window $[0,n]$ but
running the movie backwards\footnote{For simplicity we assume that
the kinematical time-reversal $\pi$ is identity, or that the
variable $X$ is even under time-reversal.}, i.e., in terms of $Y_t
\equiv X_{n-t}$. Obviously, the time-reversed process is also
Markovian and with the same stationary probability distribution
$\rho$:
\[
\Prob[Y_t = y] = \Prob[X_{n-t} = y] = \rho(y)
\]
Its conditional probabilities are obtained (not: defined) from Bayes' rule
\begin{align}\label{bayes}
\Prob[Y_t =y' | Y_{t-1}=y] &=
 \Prob[X_{n-t+1} =y | X_{n-t}=y'] \frac{\Prob[X_{n-t}=y']}{\Prob[X_{n-t+1}=y]}\nonumber\\
&= p(y',y)\, \frac{\rho(y')}{\rho(y)}
\end{align}
As a consequence, the reversed movie will be subject to the same statistical
law whenever and only if the transition probabilities for the time-reversed process
above
\[
p(y',y)\, \frac{\rho(y')}{\rho(y)} = p(y,y')
\]
equal that of the original process (right-hand side).
That is certainly the case when the transition probabilities take the form
\begin{equation}\label{db}
p(y,y') = \Phi(y,y')\,e^{[V(y) - V(y')]/2},\quad \Phi(y,y')=\Phi(y',y)
\end{equation}
for which
$\rho(y) = e^{-V(y)}/Z$ is stationary\footnote{$Z$ is a normalization.}.
The condition \eqref{db} is referred to as that
of {\it detailed balance}.\\

Such a scenario can be realized for a reduced description starting
either from classical or from quantum mechanics. If we consider a
classical Hamiltonian system $H = \frac{p^2}{2m} + V(q)$ on the
constant energy surface, there is a natural and invariant measure:
the so called Liouville volume-element measuring the phase space
volume $|M|$ of a region $M$.  Supposing of course that the energy
is an even function of the momenta, $|\pi M| = |M|$ as there are
as many states with positive or with negative momenta on the
energy surface.\\

Let us now select two regions $A$ and $B$ in phase space. They
could for example select all states with a particular density- or
velocity-profile. With respect to the Liouville measure, we may
ask what is the probability to find the state in the region $B$ at
time $t$ when, at time $t=0$, the state was in the region $A$. The
formula is
\[
\Prob[y_t \in B| y_0 \in A] = \frac{|A\cap f_t^{-1}B|}{|A|}
\]
giving the volume-fraction of states in $A$ that evolve to $B$. In
the same way, replacing $A$ with $\pi B$ and $B$ with $\pi A$,
\[
\Prob[y_t \in \pi A| y_0 \in \pi B] = \frac{|\pi B\cap
f_t^{-1}\pi A|}{|\pi B|} = \frac{|B\cap\pi f_t^{-1}\pi A|}{| B|}
\]
Using \eqref{eq:kineticinversalCL}, together with Liouville's
theorem $|f_t^{-1} M| = |M|$, we conclude
\begin{equation}\label{bent}
\frac{\Prob[y_t \in B| y_0 \in A]}{\Prob[y_t \in \pi
A| y_0 \in \pi B]} = e^{S(B) - S(A)}
\end{equation}
which is the condition of detailed balance \eqref{db} where the
Boltzmann (configurational) entropy $S(M) = \log|M|$ replaces the
function $V$ and $\rho$ is being played by the Liouville measure.
Indeed, when starting the classical mechanical system from
statistical equilibrium, and one observes the statistical
distribution of (some property) of the resulting trajectory, no
distinction can be made between past and future. That is a direct
consequence of (i) the mechanical reversibility and of (ii) the
stationarity of the equilibrium. In particular, thermodynamic
equilibrium can be characterized as the condition in which
thermodynamic past and thermodynamic future are
indistinguishable\footnote{And, in nonequilibrium situations,
entropy production can be seen as a measure of the
irreversibility, see \cite{TimeReversal}.}.\\

Here comes a quantum example, at least in a toy-version.
Let us consider just one quantum spin $1/2$.
 We have the states {\it up} $\ket{\uparrow}$ and {\it down}
$\ket{\downarrow}$ as basis.  We suppose the following (Schr\"odinger) time-evolution
\begin{equation}\label{spinev}
\psi_t = \frac 1{2}\Big[
\big( \ket{\uparrow} + \ket{\downarrow} \big) + e^{-i\omega t} \big(\ket{\uparrow} - \ket{\downarrow}\big)\Big]
\end{equation}
where we have chosen to start at time $t=0$ from $\psi_0= \ket{\uparrow}$.  At time
$t=\pi/\omega$ we are in the {\it down} state $\ket{\downarrow}$ and at time $t=2\pi/\omega$
we are back where we started.  That is periodic motion with period $T=2\pi/\omega$.\\

We measure the spin-magnetization, formally described in the
observable
\[
\sigma \equiv \ket{\uparrow}\bra{\uparrow}  -  \ket{\downarrow}\bra{\downarrow}
\]
Simple algebra teaches that at time $t=T/4$ we find $\sigma=\pm 1$
with equal probability $1/2$. At the same time and accordingly,
the system is projected into the state $\ket{\uparrow}$ or into
the state $\ket{\downarrow}$. Similarly, if we would have started
at time $t=0$ in the {\it down} state $\ket{\downarrow}$ and again
measure at time $t=T/4$, exactly the same outcome statistics would
occur as when started from the {\it up} state. Let us now imagine
looking at the movie of outcomes when the spin is repeatedly
measured at times $t=T/4,2T/4,3T/4,\ldots,nT/4,\ldots$. We see a
random sequence $(\sigma_1,\sigma_2,\ldots,\sigma_n,\ldots)$ of
outcomes $\sigma=+1$ or $\sigma=-1$ with stationary probability
distribution $(1/2,1/2)$. Obviously, when playing {\it that} movie
backward, we are as bored as before. The statistics of the
outcomes as seen in the time-reversed movie is identical to the
original. Indeed, the process constituted by the successive
measurement results is a time-reversible Markov
process\footnote{In fact, it is a Bernoulli process.} (with
trivial transition probabilities $p(\sigma,\sigma') = 1/2$).\\

The notion of spin in the above example is not at all to be taken
serious.  The above could as well describe the motion of a
particle in a symmetric double well separated by a large barrier
where in reality {\it up}, respectively {\it down}, stand for wave
functions\footnote{Say linear combinations of the ground state and
the first excited state.} supported (for the most part) in the
{\it right} or in the {\it left} well; instead of measuring the
spin we then speak about measuring the {\it right}/{\it left}
position of the particle\footnote{The analogy is borrowed from
Section 2.2 in \cite{dgz}.}.\\

Note that everything above has relied heavily on the presence of an
underlying stationary distribution. For the Markov process, it was the
stationary distribution $\rho$ that played an essential role in \eqref{bayes}.
For classical mechanics, it was the presence of the Liouville measure.
In the quantum example, it was coin tossing.

%==========================================================================================
\subsection*{Quantum free evolution}
%==========================================================================================

One of the most visible formal differences between the
Schr\"odinger equation
\begin{equation}\label{Se}
i\hbar \frac{\partial \psi}{\partial t} = H\psi
\end{equation}
and Newton's law is that \eqref{Se} is first order in time while
Newton's $F=ma$ is second order\footnote{One could argue that
Schr\"odinger's equation consists of two first order equations
(since $\psi$ is complex), very much analogous to Hamilton's
equations of classical mechanics. We still think there is an
essential difference but we do not wish to elaborate here on that
issue as it is not essential for the rest of the paper.}. Clearly
then, the fact that $\psi(x,t)$ is a solution of \eqref{Se} does
not imply that $\psi(x,-t)$ is also a solution and in {\it that}
sense Schr\"odinger's equation is not time-reversal invariant. We
hasten to give the standard response, that one should also complex
conjugate:  $\psi(x,t)$ is a solution if and only if
$\psi^\star(x,-t)$ is a solution. One argument comes from the
representation of the momentum $p = -i\hbar\partial_x$, where the
complex conjugation switches the sign of the momentum. One could
reply to that by noting that there is no {\it a priori} reason
that the momentum should change sign under
time-reversal.\footnote{After all, in $p = -i\hbar\partial_x$,
there is only a spatial (and no time-) derivative.} Furthermore it
is not clear in general how to realize operationally a complex
conjugation on the wave function of a system. Nevertheless, the
more fruitful response is to complement time-reversal with a
certain operation on wave functions much in the spirit of
\eqref{eq:kineticinversalCL} as we will now explain.\\

One of the advantages of the abstraction around
\eqref{eq:kineticinversalCL} is that the definition of mechanical
reversibility also applies to the free evolution of the quantum
formalism, i.e., the evolution on wave functions, say in the
position-representation, as given by the standard Schr\"{o}dinger
equation  \eqref{Se}. Following the proposal of Wigner
\cite{Wigner}, the recipe for time-reversal is to apply complex
conjugation.  More generally, the transformation $\pi$ of above is
now an anti-linear involution on Hilbert space. We get
time-reversal symmetry when that $\pi$ commutes with the quantum
Hamiltonian $H$. Since the Schr\"{o}dinger evolution is given by
$U(t) \equiv e^{-itH/\hbar}$, equation
\eqref{eq:kineticinversalCL} can now be written as
\begin{equation}
\label{eq:kineticinversalQM}
\pi U(t) \pi = U(t)^{\dag} = U(t)^{-1}
\end{equation}
where still, in a way,  $\pi(q,p) = (q,-p)$, albeit through a
different mechanism.\\

We conclude that not only the (classical) Newton's law\footnote{To be more precise,
we should specify the forces --- think of gravity.  One can also include Maxwell's
equations or Einstein's equations but that would take us too far.}
but also the Schr\"odinger equation\footnote{Or, for that matter, Dirac's
equation.  We do not wish to speak about possible time-symmetry breaking
due to the weak interaction.} are effectively invariant under dynamical
time-reversal:  for the free quantum flow, future and past are mere
conventions and can be described by the same laws.

%==========================================================================================
\subsection*{Quantum formalism}
%==========================================================================================

Since von Neumann \cite{vonNeumann}, textbook quantum mechanics
teaches us to complement the (linear) Schr\"odinger evolution by
the so called reduction or collapse of the wave function to avoid
the infamous measurement problem. The after-measurement wave
function is obtained from the wave function before measurement by
a highly nonlinear and \emph{stochastic} transformation; the
measurement is exactly the point where
statistics enter the quantum formalism.\\

The role and status of the collapse and the associated measurement
problem have been and still are extensively discussed in the
literature\footnote{... and at coffee-breaks or at lunch.}; that
is not our main task here. Most physicists prefer not to speak
about collapse of wave functions but give no alternative (or, what
is worse, confuse decoherence with collapse). We hope that they
would at least agree all the same that the collapse can be taken
as an effective description of the entire measurement process. If,
for the time being, we are happy with a pragmatic interpretation
of quantum mechanics, then standard quantum mechanics works
perfectly well and the measurement collapses the wave function
\emph{for all practical purposes}\footnote{We borrow the phrase
from John Bell's
paper \cite{agm}.}.\\

At this point, the plot thickens. Generally speaking, it is not
possible to reverse the reduction. One cannot {\it de-measure} or
{\it de-collapse} the wave function. The after-measurement wave
function is very much limited --- it must be an eigenstate of the
measured observable --- but not the before-measurement wave function.
That is the point where for some problems and for
others solutions seem to arise. Some find it odd that the rules of
the game seem to break time-reversal symmetry on a rather fundamental
level; it has less esthetic appeal and nothing of it remains
for the limiting classical mechanics. For others an opportunity
seems created to give dynamical derivations of  second law behavior.\\

Does that irreversible behavior of wave functions in the
measurement formalism means that quantum mechanics is
irreversible? Clearly the answer depends on how serious one takes
that measurement formalism, or even on what one means by quantum
mechanics. Remains that some see in it a manifestation of a
fundamental spontaneous breaking of the time-symmetry at the
beginnings of time\footnote{We refer to the quantitative analysis
of R. Penrose in Chapters 7 and 8 of \cite{Penrose}.}.

%==========================================================================================
\subsection*{Statistical reversibility II}
%==========================================================================================

The standard quantum theory with its measurement formalism is
concerned exclusively with giving predictions for frequencies of
future measurements. As a matter of logic, if one regards the
collapse as a device that works for all practical purposes but has
no ambition to be {\it fundamentally true}, there is no point to
blame it for breaking mechanical time-reversal symmetry and to
base  major theoretical consequences on that. If one works on the
level of a statistical theory, where one is happy to calculate
probabilities of outcomes, to be consistent, one should only
enquire about statistical reversibility.  The time-reversal
symmetry breaking of the collapse is then only pointing to a
(theoretical) inadequacy of the standard interpretation with no
further consequences except for giving yet another argument that
the collapse procedure {\it cannot} be dynamically deduced from
the (time-reversible and linear) Schr\"odinger evolution for
the complete system + apparatus. \\

At first sight, the quantum spin example \eqref{spinev} seems a
promising start to recover time-symmetry. Before we step back to
meditate, we see whether we can generalize it to include for
example higher dimensional projections\footnote{Other
generalizations, like considering {\it fuzzy} measurements are
of interest but will be skipped here.}.\\

On a more formal level, one considers a finite-dimensional Hilbert
space and projections $P_\alpha$ where $\alpha$ runs over a set of
linear subspaces.  At the same time, $\alpha$ refers to an outcome
of a measurement of some observable\footnote{For notational
simplicity we take in what follows measurements of the {\it same}
observable. That is however not needed.}. For example, we have a
system of $N$ distinguishable spin $\frac{1}{2}$ particles and we
look at the total magnetization in the $z$-direction
\begin{equation}
\label{eq:magnetization} m_z \equiv \frac{1}{N} \sum_{i=1}^N
\sigma^z_i
\end{equation}
There are $N+1$ outcomes $\{\alpha\}$ for a measurement of $m_z$.
with corresponding orthogonal eigenspaces and projections
$P_{\alpha}$ of different dimensionality $d_{\alpha} =
\Tr[P_{\alpha}]$. The sum $\sum d_\alpha=d$ is the dimension of
the Hilbert space. We refer to the $\alpha$'s as {\it conditions}.
The time evolution on the spin-system is described by a
Hamiltonian $H$, implemented by a unitary $U(t) = e^{-itH}$
($\hbar$ equals {\it one}).\\

We start the system  in condition $\alpha$ with probability $d_\alpha/d$.
That means that the initial density matrix is
\[
\rho = \sum_{\{\alpha\}}\frac{d_\alpha}{d} \frac{P_\alpha}{d_\alpha} = 1/d
\]
the normalized unit matrix. We measure the magnetization at fixed
times\footnote{Choosing unequal time-intervals between successive
measurements makes no difference.} $t=0,1,2,\ldots$ and we ask for
the probability to find then the system consecutively in
conditions $\alpha_0,\alpha_1,\alpha_2,\ldots$. Writing for short
the sequence of outcomes $\omega = (\alpha_0 , \alpha_1, \ldots,
\alpha_t)$, that probability equals
\begin{equation}\label{probo}
\Prob[\omega] = \frac{1}{d}
\Tr[ \, U P_{\alpha_{t-1}} \ldots U P_{\alpha_1} U
P_{\alpha_0} U^{\dag} P_{\alpha_1} U^{\dag} \ldots
P_{\alpha_{t-1}} U^{\dag} \, \, \, P_{\alpha_t}]
\end{equation}
where $U \equiv U(t=1) \equiv e^{-iH}$.\\

Again one can look at the time-reversed sequence. Like in the
classical case, not only do we have to reverse the order of the
conditions but we should also replace each projection $P_\alpha$
by its kinematical time-reversal $\pi P_\alpha \pi = P_{\alpha'}$.
By that last procedure every condition $\alpha$ has a counterpart
$\alpha'$.  We now ask about the probability to measure, in the
indicated order, and starting from the same density matrix
$\rho=1/d$, the conditions
\begin{displaymath}
\Theta \omega = (\alpha_t' , \alpha_{t-1}', \ldots, \alpha_0')
\end{displaymath}
Using \eqref{eq:kineticinversalQM} that probability equals
\begin{equation}
\label{probto} \Prob[\Theta\omega] = \frac{1}{d} \Tr[ \, U^{\dag}
P_{\alpha_{1}}  \ldots U^{\dag} P_{\alpha_{t-1}}  U^{\dag}
P_{\alpha_t}  U P_{\alpha_{t-1}} U \ldots P_{\alpha_{1}}  U \, \,
\, P_{\alpha_0}]
\end{equation}
Upon inspection
\[
\Prob[\omega] = \Prob[\Theta\omega]
\]

In contrast with the quantum example \eqref{spinev}, the statistics
of the repeated measurements is in general no longer described by
a Markov process but it does satisfy time-reversibility: from the
statistics of outcomes we will not be able to decide whether the
movie ran forward or backward.  Note however again that we have
here a stationary situation; at every moment the probability
of condition $\alpha$ is $d_\alpha/d$.\\

If instead of looking at the joint multi-time probability one
considers the transition probabilities, one easily checks as done
in \cite{QuantumEntropy} that a condition similar to detailed
balance of \eqref{db} holds true. That is very much identical to
what was obtained in \eqref{bent} for the classical dynamics but
with the Liouville volumes now being replaced by dimensions and
the classical entropy replaced by $S(\alpha) = \log d_{\alpha}$,
the quantum Boltzmann entropy associated to the condition
$\alpha$:
\begin{equation}
\label{eq:dimensions}
\frac{\Prob[\omega|\alpha_0]}{\Prob[\Theta\omega|\alpha_t']} =
\frac{d_{\alpha_t}}{d_{\alpha_0}} \equiv e^{S(\alpha_t) -
S(\alpha_0)}
\end{equation}

One could wonder in the above (as in classical mechanics) what is
the role of the stationary density matrix chosen to be a
normalized multiple of unity. Moreover, that seems to prevent
extensions to infinite dimensional Hilbert spaces. That problem
can be avoided by considering the ensemble where one conditions on
the results of the initial and final measurements. One then asks
for the probability of outcomes
$(\alpha_1,\alpha_2,\ldots,\alpha_{t-1})$ at consecutive times
{\it given} the outcome $\alpha_0$ at time $0$ and the outcome
$\alpha_t$ at time $t$ {\it and given} that there were
measurements at the intermediate times. That is
\begin{equation}\label{abl}
\Prob[\alpha_1\ldots,\alpha_{t-1}\,|\,\alpha_0,\alpha_t]
= \frac{\Prob[\omega]}{\sum_{\alpha_1,\ldots,\alpha_{t-1}}\Prob[\omega]}
\end{equation}
Again it is easy to see that \eqref{abl} is manifestly
time-symmetric\footnote{Conditioning on past and future events is not so unphysical as one
 could imagine at first.  In various experimental situations one selects
 the sample upon verifying both initial and final states.}.
That observation was first made by Aharonov, Bergmann and Lebowitz
in 1964 \cite{ABL}. We repeat that the reversibility as in
\eqref{abl} is for the conditional probabilities of the results of
a sequence of measurements, given the results of the initial and
final measurements. That is entirely compatible with the
irreversibility in the behavior of wave functions in the
measurement formalism.

%==========================================================================================
\subsection*{Retrodiction}
%==========================================================================================

While we learn in school that prediction is difficult, especially
of the future, and retrodiction is more easy, especially of the
past, interesting questions extend in both directions.  Textbooks
in quantum mechanics usually emphasize the problem of prediction.
Conventional quantum mechanics does not think of upcoming
measurements as determining the state of the system now; the only
information that is produced is about the state of the system
subsequent to the execution of the measurement. In that sense the
theory has a time-asymmetrical twist. Nevertheless the issue of
retrodiction in quantum mechanics is not philosophical and it
appears for example in problems of quantum optics and
cryptography. From the point of view of textbook quantum
mechanics, strange effects can appear. Here is one example.\\

Suppose a spin $1/2-$particle is initially prepared at time $t_1$
with spin pointing {\it up} in the $x-$direction.  For
time-evolution we take the trivial one, with Hamiltonian zero, so
that the quantum state is unchanged through the Schr\"{o}dinger
evolution.  At a later time $t_2>t_1$ we measure its $z-$component
to find it, suppose, {\it up} in the $z$-direction. Then we are
sure that in the intermediate times $t\in (t_1,t_2)$ the spin was
also {\it up} in the $z-$direction. By the latter we mean that,
{\it if} we had measured the $z-$component, it would certainly
have been {\it up}: a trivial dynamics would not be able to
transform a {\it down} measurement into the {\it up} measurement
at time $t_2$. This is an example of the time-symmetry when fixing
initial and final results as explained at the end of the previous
section, and as contained in \cite{ABL}. Analogously, had we
measured the $x-$component of the spin at these intermediate
times, it would also certainly have been {\it up}: the trivial
dynamics retains the initially prepared state! As a consequence,
at an intermediate time, both the $x$- and the $z$-component seem
well defined as their measurements would have been unambiguous.
That statement is already somewhat strange for quantum mechanics,
but it is very strange if you would think it could only be made
{\it because} of the knowledge of the result of the later
measurement, see also \cite{aa}\footnote{Less trivial and more
sensational aspects of retrodiction in quantum mechanics are
discussed in \cite{mer,sb,vaa}.}.\\

Let us now take two distinguishable spin $1/2$ particles, again
with the trivial dynamics and initially prepared in some state
\[
\psi_0 = c_{\uparrow\uparrow}\ket{\uparrow\uparrow} +
c_{\uparrow\downarrow}\ket{\uparrow\downarrow} +
c_{\downarrow\uparrow}\ket{\downarrow\uparrow} +
c_{\downarrow\downarrow}\ket{\downarrow\downarrow}
\]
At time $t_1$ we measure the total magnetization $\sigma^z_1 +
\sigma^z_2$ in the $z$-direction and we suppose that we find
$\sigma^z_1 + \sigma^z_2 = 0$. At time $t_2>t_1$ we measure
$\sigma^z_1 - \sigma^z_2$ and the probability that it is 2 (given
the specified past) is equal to\footnote{One should use the matrix
$s^z = \frac{\hbar}{2} \sigma^z$ as the spin-observable. To ease
the notation, we use $\sigma^z$ instead.}
\[
\frac{|c_{\uparrow\downarrow}|^2}{|c_{\uparrow\downarrow}|^2 + |c_{\downarrow\uparrow}|^2}
\]
We now time-reverse.  Assume that at time $t_2$ we in fact measure
$\sigma^z_1 - \sigma^z_2 = 2$. That corresponds to the state
$\ket{\uparrow\downarrow}$, which now serves as an initial state
for the reversed dynamics. Then, under this reversed dynamics,
certainly the magnetization $\sigma^z_1 + \sigma^z_2 = 0$ when measured at time $t_1<t_2$.\\

Hence, the probability for measuring $\sigma^z_1-\sigma^z_2=2$ at
time $t_2$ given zero magnetization at time $t_1$ and started from
$\psi$ at time $t<t_1<t_2$ depends on $\psi$ through the
coefficients $c_{ij}$, and can for example be equal to
$\frac{1}{2}$. At the same time the probability to measure
$\sigma^z_1 + \sigma^z_2 = 0$ at time $t_1$ before one sees
$\sigma^z_1-\sigma^z_2=2$ at time $t_2$ just equals
one\footnote{We think of that as the example on page 357
of Penrose's book \cite{Penrose} translated to spin-language.}!\\

All that is not in contradiction with the statistical
reversibility of the previous section. It only shows that even on
a microscopic level the dimensionality of the eigenspaces of the
various observables involved in the art of retrodiction, plays a
crucial role. Observe that when all $c_{ij}$ are chosen equal, the
ratio of forward to backward probabilities equals 1/2, precisely
the ratio of corresponding eigenspace dimensions as in the
detailed balance condition \eqref{eq:dimensions}. There are
classical analogues of the example but not on the microscopic
level; one must go to stochastic or reduced descriptions as for
example in the understanding of thermodynamic irreversibility.

%==========================================================================================
\subsection*{Thermodynamic irreversibility}
%==========================================================================================

Dynamical or statistical reversibility does not exclude
macroscopic or thermodynamic irreversibility.  That is relevant
because it says we do not need explicit breaking of time-reversal
invariance of the microscopic laws to account for the observed
thermodynamic irreversibility.  In fact, it is largely unimportant
whether the microscopic time-evolution is reversible or not:
macroscopic irreversibility typically obtains. Moreover, the
mechanism that leads to second law behavior or entropy increase is
quite independent of the classical or quantum nature of the system.\\

Thermodynamic irreversibility is a statement about the
\emph{typical} temporal behavior of macroscopic observables.
Something quite remarkable can happen: as the number $N$ of
degrees of freedom of the system gets very large and when the
system is observed over the appropriate time scales, these
macroscopic observables can start to follow an autonomous
evolution. Usually one considers density or velocity profiles
satisfying the kinetic or the hydrodynamic equations that
constitute the phenomenology of time-dependent thermodynamics.
Moreover, these equations are not only true {\it on average}; they
are {\it typically} true, in the sense of the law of large numbers
applied to the initial condition: the spreading or variance around
the observable should go to zero for large $N$, while the
expectation value solves a first order differential equation with
a given initial value.\\

Obviously, to actually prove the emergence of autonomous
macroscopic behavior with relaxation from far-from-equilibrium to
equilibrium is mostly beyond our abilities today. However, that
should not distract us from the physical mechanism behind
thermodynamic irreversibility: as $N \rightarrow \infty$, the
phase space volumes (classical systems) or the dimensions of the
subspaces (quantum systems) corresponding to a particular
macrostate, become overwhelmingly different. Conspiracies and
special initial conditions aside, the dynamics brings the system
to a region in phase space where the Boltzmann entropy is larger,
precisely due to this huge difference in scales; a conclusion one
draws purely by enumerating and counting the allowed
possibilities.
% Because of this huge difference in scales, except for
% conspiracies, the dynamics brings the system to a region
% in phase space where the Boltzmann entropy is larger, a
% conclusion one draws purely by enumerating and counting
% the allowed possibilities.
In the classical formalism, this set-up also explains why the so
called Poincar\'{e} recurrences are irrelevant when studying
relaxational behavior. In the quantum formalism, the same is true
for the quasi-periodicity that occurs in finite
Hamiltonian systems.\\

In sufficiently simple set-ups we do succeed in proving autonomous
relaxation equations for the macroscopic variables and we thereby
confirm the general picture above.  An example is treated in full
detail in \cite{QuantumKac} as a quantum extension of a model that
was first conceived by Mark Kac \cite[p.\,99]{Kac}. That quantum
Kac model shows irreversible behavior for the magnetization as a
function of time and the corresponding entropy in the model
increases in time towards its equilibrium value.  Yet the dynamics
is given by standard quantum mechanics for a finite system of
spins.

%==========================================================================================
\subsection*{Conclusions}
%==========================================================================================

1.  If one includes the collapse of the wave function in the
mechanics of the quantum world, then there is no reversibility
even on the microscopic level.  One cannot demeasure things.  One
can however associate fundamental importance to that kind of
microscopic irreversibility only up to the point where one considers
the collapse as a fundamental input of quantum mechanics rather than
as an effective device {\it for all practical purposes}.
\\

2.  If one understands the collapse of the wave function within
the standard statistical interpretation of quantum mechanics it is
appropriate to ask for statistical reversibility, i.e., in terms
of probabilities of histories. In that case it is quite similar to
the situation of {\it microscopic reversibility} or {\it detailed
balance} for transition probabilities as obtained also classically
from Hamiltonian dynamics.  Statistical reversibility
is satisfied within standard quantum mechanics.\\

3. Thermodynamic irreversibility is an emerging property in
macroscopic behavior for which the reasons are basically unchanged
in the transition from classical to quantum dynamics. In
particular, such macroscopic irreversibility can be expected and
sometimes is obtained on appropriate time-scales for quantum
unitary evolutions with respect to typical initial data.\\

As a final remark, we like to add that a fully mechanically
reversible version of quantum mechanics exists which is, for all
we know, empirically equivalent with standard quantum mechanics:
the Bohmian equations of motion\footnote{They consist of an
equation for the wave function, nothing else than the
Schr\"odinger equation, complemented with an equation for the
positions of all particles. We refer to \cite{aa} for a discussion
on the issue of retrodiction in Bohm's theory.}. However one does
not necessarily need to resort to a modification of standard
quantum practice; in a practical sense standard quantum theory can
account for both the macroscopic irreversibility and statistical
reversibility, as present in our daily experience.

\newpage
% ===================================================================
% BIBLIOGRAPHY
% ===================================================================

{%
  \let\section=\subsection

}

\begin{thebibliography}{100}
{\small

\bibitem{aa}
Y. Aharonov, D. Albert, {\em The Issue of Retrodiction in Bohm's Theory},
in: {\it Quantum Implications}, B.J. Hiley and F.D. Peat (eds.),
Routledge \& Kegan Paul, New York, 224--226 (1987).

\bibitem{ABL}
Y. Aharonov, P.G. Bergmann, J.L. Lebowitz, {\em Time symmetry in the
quantum process of measurement},
Phys. Rev. B \textbf{134}, 1410--1416 (1964).

\bibitem{agm}
J.S. Bell, {\em Against measurement}, Physics World \textbf{8}, 33--40 (1990).

\bibitem{QuantumEntropy}
I. Callens, W. De Roeck, T. Jacobs, C. Maes, K. Neto\v{c}n\'{y},
{\em Quantum entropy production as a measure for irreversibility},
Physica D \textbf{187}, 383--391 (2004).

\bibitem{deHaan}
M. de Haan, {\em Vers l'historicit\'{e} en th\'{e}orie quantique},
Physicalia \textbf{26}, 29--31 (2004).

\bibitem{QuantumKac}
W. De Roeck, T. Jacobs, C. Maes, K. Neto\v{c}n\'{y}, {\em An Extension of the Kac ring model},
J. Phys. A: Math Gen. \textbf{36}, 11547--11559 (2003).

\bibitem{dgz}
D. D\"urr, S. Goldstein, N. Zanghi, {\em Quantum Chaos, Classical Randomness,
and Bohmian Mechanics}, J. Stat. Phys. {\bf 68}, 259--270 (1992).

\bibitem{Kac}
M. Kac, {\em Probability and related topics in physical sciences}
(chapter 14), Interscience Publishers Inc., New York (1959).

\bibitem{Snipers}
P. Schewe, B. Stein, {\em Reversing time to catch snipers},
Physics News Update \textbf{687}, \texttt{http://www.aip.org/pnu/2004/split/687-1.html}

\bibitem{TimeReversal}
C. Maes, K. Neto\v{c}n\'{y}, {\em Time-reversal and entropy},
J. Stat. Phys. \textbf{110}, 269--310 (2003).

\bibitem{mer}
D. Mermin, {\em Limits to Quantum Mechanics as a Source of Magic Tricks: Retrodiction
and the Bell-Kochen-Specker Theorem}, Phys. Rev. Lett. \textbf{74}, 831--834 (1995).

\bibitem{Pauli}
W. Pauli, {\em \"{U}ber das H-Theorem vom Anwachsen der Entropie vom Standpunkt der neueren Quantenmechanik},
in ``Probleme der modernen Physik'', Sommerfeld Festschrift, Leipzig,  30--45 (1928).
% Festschrift = Arnold Sommerfeld zum 60. Geburtstage gewidmet von seinen Schülern

\bibitem{Penrose}
R. Penrose, {\em The Emperor's New Mind}, Oxford University Press, Oxford (1989).

\bibitem{sb}
S. Ben-Menahem, {\em Spin-measurement retrodiction}, Phys. Rev A \textbf{39}, 1621--1627 (1989).

\bibitem{vaa}
L. Vaidman, Y. Ahoronov, D.Z. Albert, {\em How to ascertain the values of $\sigma_x$,
$\sigma_y$, and $\sigma_z$ of a spin-$1/2$ particle}, Phys. Rev. Lett. \textbf{58}, 1385--1387 (1987).

\bibitem{vonNeumann}
J. von Neumann, {\em Mathematical Foundations of Quantum Mechanics} (translated by R.T. Beyer),
Princeton University Press, Princeton (1955).

\bibitem{Wigner}
E. Wigner, {\em \"{U}ber die Operation der Zeitumkehr in der Quantenmechanik},
Nachr. Akad. Ges. Wiss. G\"ottingen \textbf{31}, 546--559 (1932).
}
\end{thebibliography}
\end{document}